# IT Students Career Confidence and Career Identity During COVID-19

*Sophie McKenzie*

School of Information Technology

*Deakin University, Geelong, Australia, 3216*

Background: COVID-19 disrupted the professional preparation of university students, with less opportunity to engage in professional practice due to a reduced employment market. Little is known about how this period impacted upon the career confidence and career identity of university students.

Research Question: This full research paper explores the career confidence and identity of university students in Information Technology (IT) prior and during the COVID-19 period.

Methods: Using a survey method and quantitative analysis, ANOVA and Kruskal-Wallis tests with different sensitivity and variance standards were used during analysis to present mean and mean rank of data collected during 2018, 2019, 2020 and 2021. 1349 IT students from an Australian University reported their career confidence.

Findings: The results indicate IT students' career confidence maintained during the period. In 2021, the results indicate increased career commitment of IT students showing higher professional expectations to work in IT along with greater self-awareness regarding their professional development needs.

Implications: Even with increased career confidence as observed in this study, supporting university students to explore their career options and build upon their career identity, and more broadly their employability, remains an important activity for universities to curate in their graduates.

## 1   INTRODUCTION

The period between 2020 and 2022 has become known as the time of the COVID-19 pandemic, where the everyday activities of societies around the world were disrupted. Working from home became expected with many professionals experiencing major changes in their working life. For those studying at university, remote learning dictated all learning experiences. In addition, the opportunity to engage in professional practice experiences on-site, such as work experience or a placement/internship, were limited. The impact of this time on students' career confidence, career identity and professional outcomes is not yet known. Career confidence is a student's perceived ability to pursue a career, with career identity demonstrated through a student's motivation, interests and competencies directed towards acceptable career roles. Individuals need to develop a career identity to navigate the social and work-related insecurities of modern society. As the world moves into a post-pandemic phase, it is important to understand how this period impacted the professional preparation of university students. Without suitable preparation, students may not be best prepared for their career, nor have a suitable level of employability ("the ability to find, create and sustain meaningful work across the career lifespan and in multiple contexts" (Bennett, 2016)). To assist, this study explored the career confidence and career identity of students in Information Technology (IT) prior and during the COVID-19 period.

Information Technology or IT is a broad professional field that encompasses a variety of roles across diverse sectors of the industry. Generally, when working in IT occupations may include a programmer or developer, technology consultant, data analyst, application support analyst, or web designer/ developer. The rapid onset of digital skills from COVID-19 has emphasised the importance of IT skills across occupations (National Skills Commission 2021). In addition, in Australia up to 2026, there is a projected growth of 16.8% in professional and technical roles, where roles in IT are often found (Australian Government, 2023). When it comes to working in IT students have realistic aspirations of what job they would like to do, despite broad professional options (McKenzie et al., 2017). Digital skills and working in IT remains a growing and realistic occupation for many.

## 1.1 Problem

To understand students' future professional performance, this study explored the career confidence of IT students while at university. Over 4 years, 1349 students from an Australian University reported their career confidence. Specifically, this paper will address the following research questions: What is the career confidence of students in ICT, leading up to and during the COVID-19 period? Further, using the understanding of students' career confidence during the COVID-19 period, this paper will explore, has COVID-19 changed IT students' career identity?

**1.2 Review of Relevant Scholarship**

A career is where a person engages in an occupation or professional work for a significant period of their life. A career is shaped by a person's beliefs, behaviours, and a variety of personal, interpersonal, societal, and environmental factors (Hall, 2002; Patton & McMahon, 2014). While at university it has become reasonable to expect that students will participate in and expand upon their career expectations, interest and goals through their formal study and other career supportive experiences, such as participating in professionally situated learning experiences. Students are now expected to regularly engage in and reflect on their career progress during their study to build their career prospects but to also more broadly build their employability and adaptability to have work-force participation. It has been noted for some time that there is a link between career development and employability, with the concepts going together in supporting an individual in their working life (Watts, 2006). Employability is an amorphous term, defined in a variety of ways and can be conceived differently depending on economic context and discipline. A process of career development is considered important to enhance a student's ability for longer-term employability (or sustainable employability) (Watts, 2006).

*1.2.1 What is Employability?*

\* defines employability in higher education as: "the ability to find, create and sustain meaningful work across the career lifespan and in multiple contexts" (Bennett, 2016). \* emphasizes the importance of employability thinking, in that students' need their cognitive and social development supported to ensure they are capable and "informed individuals, professionals and social citizens" (Bennett, 2016). To support students' employability explicit career- and life-long identity and employability work need to be incorporated in the higher education curriculum. A process of career development can contextualise employability needs and align skills, attributes and interest with career choice (Jollands, 2015; Nagarajan & Edwards, 2008; Qenani et al., 2014). (Dacre Pool & Sewell, 2007) and (Watts, 2006) proposed a crossover between models of career development and employability, demonstrating how skills development goes hand-in-hand with informing career choice.

In their study, Römgens, Scoupe and Beausaert define employability as "the prevalence of disciplinary knowledge and general skills needed to perform a given occupation; the importance of social skills, including networking skills; the necessity to continue learning and to (pro-) actively as well as passively adapt to changing situations and environments;

and meta-cognitive skills for reflecting on one's goals and values, ambitions and identity. Both research streams complement each other in explicating the importance of respectively emotional regulation (from higher education) and achieving a healthy work-life balance (from workplace learning)" (Römgens et al., 2020, p. 2589).

Supporting students' broader employability is an important consideration in Australia. In 2020, the Australian Government announced the Job-ready Graduates Package. The National Priorities and Industry Linkage Fund (NPILF), introduced under the package, will allocate block grants to universities to support enhanced engagement with universities and industry to produce job-ready graduates (Australian Government Department of Education, 2022). Student engagement with industry through work-integrated learning forms a core part of the metric, supporting the need to support students' preparation for employment, and more broadly, their employability. Employability support will translate not only to their success while at university, but to longer term career persistence, career confidence and labour market participation.

*1.2.2 Career Confidence*

Career confidence is a student's perceived ability to pursue a career (Caza et al., 2015) with confidence growing throughout a person's life. It is important for students to feel confident about their career choices, with increased self-efficacy (i.e., confidence in one's ability to accomplish specific goals (Bandura, 1986)) a key factor in informing career choice. Bandura argues that career development is socially mediated, with social cognitive career theory informing how students form their career confidence and choice. Figure 1 describes a model of career development, building off SCCT, that encourages IT students career identity, interest and career choice (McKenzie et al., 2021).

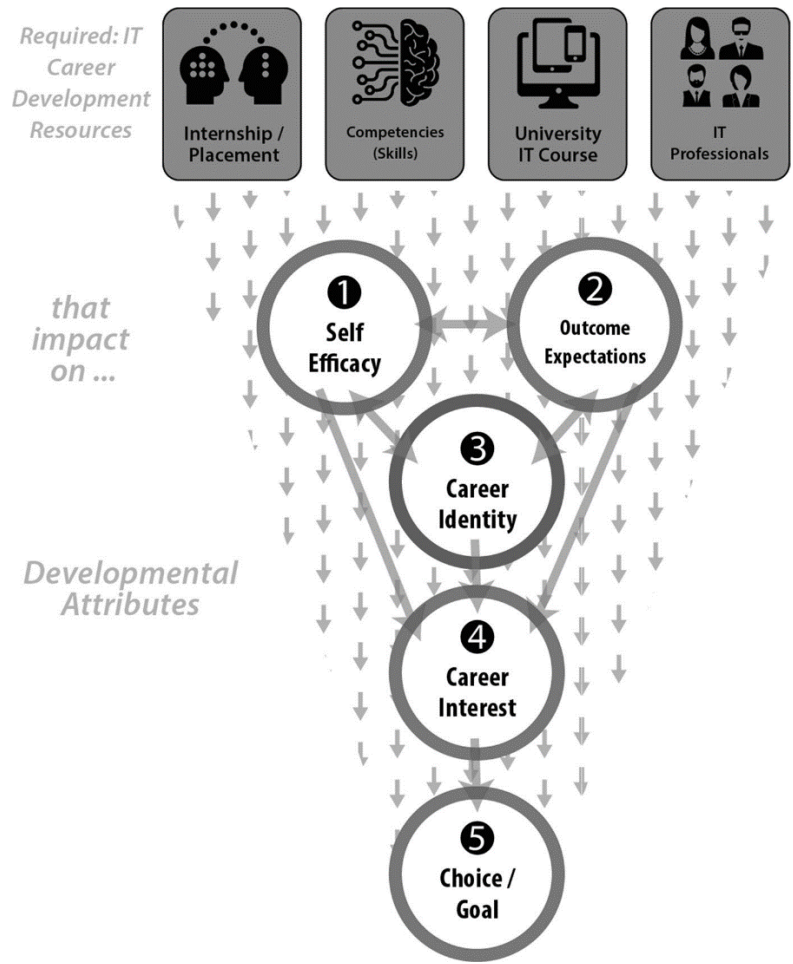

Figure 1 Model of Career Development for IT Students

This model deviates from the original SCCT model through addition of career identity as a developmental attribute. *, * and * (McKenzie et al., 2021) found that for IT students' formation of a pre-professional or career identity contributes to developing career interest and goals. Seeing themselves in the profession impacts upon IT students career interest, with career interest a mediator to achievement of career goals (McKenzie et al., 2021). In this study, career confidence relates to student's decisional self-efficacy, career identity, program relevance/interest, and goal setting (ref choice of major paper). Career identity for IT students takes an important role in informing the development of their career goals.

*1.2.3 Career Identity*

Erikson (1994) indicated that a person's identity is developed via eight psychosocial stages, from identity's initial formation through to subsequent evolution in adulthood. Identity development occurs for every individual, and is informed by biological, physiological and social variables. In addition, (Mahler, 2008) argued that mental and physical health, intelligence, culture, gender, ethnicity, religious and spiritual beliefs, geographic and economic factors, and learning and work situation also impact on identity development. Of significance in this research is the sub-category of identity, which focuses on career. The work a person does throughout their life informs their development of their career identity (Hall 2002). Meijers (1998, p. 1) defines career identity as "a structure of meanings in which the individual links their own motivation, interests and competencies with acceptable career roles". During their education a student will expand their identity to evolve into a graduate or professional identity, contributing to their career identity (Nadelson et al., 2015; Smith et al., 2019). Nadelson et al., (2015, p. 705) described professional identity as "the attributes, skills, knowledge, beliefs, practices, and principles, which are representative of professionals within a profession". Students' perceptions of their professional selves are developed internally and validated externally during their time at university (Holmes, 2001; Nadelson et al., 2015). Meijers (1998) argued that individuals need to develop a career identity to navigate the social and work-related insecurities of modern society.

Scott and Ciani (2008) refer to vocational identity rather than career identity, arguing that adolescents need to develop their vocational identity to assist occupational awareness and career exploration. A career identity is informed by a person's traits and characteristics as well as internal aspects such as values and beliefs (Mahler, 2008; Super, 1980). As a psychosocial process, career identity is constantly developing, informed through meaning making and experience. Hall (2002) argued that for success in a 21st century career, an individual must be adaptable and pass through multiple work role transitions. In each transition they learn from their experiences and add to their career identity. Smith et al., (2019) found that university students were more confident in expressing aspects of a professional identity when they engaged in experiential learning opportunities, such as work experience.

Holmes (2001) outlined a model for understanding graduate identity. The model shows that an individual engages in self-identification/claim of their career/work identity, which is then affirmed by others (such as a recruiter). Identity may be disaffirmed by the graduate, or others, resulting in the need for identity development/adjustment. The model describes different modalities of emergent identity that arise from the interaction been claiming/disclaiming identity status and affirmation/disaffirmation by significant others. Selection of a suitable and satisfying work environment occurs when an individual has an informed and affirmed sense of identity.

Gianakos (1999) described students' career identity through four patterns of career choice: stable through adolescence and adulthood, conventional through adolescence with a change during adulthood, multiple trial through adolescence and adulthood but onto a second career, and unstable through adolescence and adulthood. These four patterns can be categorised into two groups: students who have decided on a career choice (stable and multiple trial) and those who are undecided (conventional and unstable). Undecided students have been found to be anxious, externally controlled, and confused as to their identity (Gianakos, 1999). Compared to their decided peers, undecided students use career decision-making styles that are less cognitively complex and adaptive (Harren, 1979) but are more emotional and impulsive, thereby interfering with career exploratory behaviours (Blustein et al., 1989). Gianakos (1999) also found that increased levels of self-efficacy were exhibited in the career decided group. Individuals whose career choice histories reflect unstable or uncommitted patterns are likely to have lowered beliefs in career decision-making self-efficacy (Gianakos, 1999). To make best use of career development resources available to them during university, and to inform their career choice, students need to have a strong sense of their career identity. * et al., (2021) demonstrated that for IT students that SCCT helps to understand the factors that impact on students' career development, with their prior studies in IT and their need for access to "IT professionals" informing their career interest and identity. IT students have a strong intrinsic interest to study IT (McKenzie & Bennett, 2022), and rely on professional experiences (such as a placement or internship) to inform their career identity (McKenzie et al., 2021). However, the impact of COVID-19 on students forming career confidence and identity is unknown. Through a sample of IT students, this study will explore what impact COVID-19 has had on the professional prospects of IT students.

### 1.3 Study Context: IT students in Higher Education in Australia

Information Technology (IT), Information Communication Technology (ICT) or Computing are umbrella terms often used to describe one broad professional area, loosely defined as work that includes computer system design and telecommunications and internet services (Palmer et al., 2018). Typical job roles that define an IT professional include manager, project manager, business and systems analyst, software engineer, computing networking professional, full stack developer, software and application programmer, sales professional, database and systems administrator, multimedia specialist, and support and test engineers (Palmer et al., 2018). Professions considered 'outside' of IT now involve a number of activities that are considered technical, showing diversity in the ways in which IT is integrated into the working world. IT (or ICT) education has been a formal program in higher education institutions the world over since the 1980s. Globally, IT education can be defined as computing engineering, computer science, information systems, or software engineering (Palmer et al., 2018; SFIA Foundation, 2021). IT education now addresses many 21st century technology innovations such as: Mobile application development, artificial intelligence, data science, games development and cyber security.

In the past 10 years the graduate employment landscape in Australia has changed considerably. COVID-19 presented a number of challenges for the workforce, however recent data shows a recovery in employment outcomes and prospects across a large part of the Australian workforce (National Skills Commission, 2022). In the 2021 employment outcomes update (National Skills Commission, 2022) labour market growth shows that over the 2020-2021 period 92% of total employment growth has been in occupations where some level of post-secondary school qualifications is usually required (Australian Government, 2023). In IT, the role of a software and applications programmer is projected to have growth of 27% leading up to 2026, with database and systems administrators, and cyber security specialists predicted to grow almost 40% (National Skills Commission, 2022). * et al. (2018) defined the job market for Australian IT bachelor graduates as dynamic and competitive, with strong competition from the global IT workforce, yet new graduates are finding it increasingly difficult to secure initial employment in the IT industry (Palmer et al., 2018).

In this study, one major provider of higher education has been selected as the source for gathering career confidence of IT students. Deakin University is a major provider of distance and online education, and also teaches on-campus at four campuses located in three cities in the State of Victoria, spanning metropolitan, regional and rural locations. Deakin University currently teaches on a trimester system, with three teaching periods per year of equal duration and status. The discipline of Information Technology rests within the Faculty of Science, Engineering and Built Environment. The School of Information Technology (SIT) offers undergraduate and postgraduate IT programs. The courses offered specialize in computer science, software engineering, networking, artificial intelligence, data analytics, cyber security and creative technology (games development, web and mobile applications). These programs are delivered in both on-campus and off-campus modes, with a focus on an industry and professional society informed curriculum, to build IT graduates. Deakin University aligns their course design to the Australian Computer Societies (ACS) 'core body of knowledge' skills framework. To build graduate employability, career education has been provided to IT students during the period of this study through the graduate employment division, with career education scaffolded and embedded in the curriculum in 2018.

Deakin University was chosen as the site of this study because of its approach to supporting graduate employability. The active engagement of students with career development activities enabled the student experience to be explored across the IT undergraduate program. In addition, IT graduates from Deakin University are major contributors to the local and global economy, showing skills constructed during their time at university meet industry requirements. The student profile of Deakin University is like other universities in the region, enabling the student experience to be generalised to other institutions. This study investigated; What is the career confidence of students in IT, leading up to and during the COVID-19 period? Further, using the understanding of students' career confidence during the COVID-19 period, this paper will explore, has COVID-19 changed IT students' career identity?

## 2 METHOD

### 2.1 Participant Characteristics and Sampling Procedures

In 2018, 2019, 2020 and 2021 n= 1349 students in a second year IT subject at Deakin University completed the developing employability survey as a part of their developmental learning related to being an IT professional. Ethical clearance was obtained from the human research ethics committee from the first author's institution.

### 2.2 Data Collection

The developing employability survey asked students to complete an extensive list of questions regarding their self-perceived confidence on a range of scales that relate to employability, with over 200 questions posed to each student.

Students responded to Likert-style items to assess their confidence in relation to self-management, career decision-making, self-esteem, academic self-efficacy, identity construction, the citizen-self, emotional intelligence, and perceived learner and graduate attributes. Students provided demographic information as a part of their survey response. Students could also respond to optional open questions relating to their work and study backgrounds, career intentions, choice of major and their feedback about their current courses (programs). The developing employability survey has been validated as a scale to identify students current employability development (Bennett & Ananthram, 2022).

Based on their responses to the developing employability survey, each student received an employability report showing their employability outcomes on seven literacies: core, career, learning, rhetorical, emotional, ethical, and digital. For this study and drawn out of the seven literacy groups, students' career confidence and identify is reported against the factors of: self-awareness related to program of study, career commitment, career exploration and awareness and goal directed behaviour. These factors were extracted from the broader developing employability student survey due to their focus on career related components such as decisional self-efficacy (self-awareness), career identity (career commitment), career identity (career commitment), program relevance (interest), and goal setting (goal directed behaviour) (McKenzie & Bennett, 2022).

Self-awareness related to program of study is a social cognitive construct of motivation that is understood to positively influence engagement and self-regulation (Bandura, 1986). Self-awareness related to program of study concerns students' perceptions of how their study program relates to their future career. In the developing employability survey items for self-awareness related to program of study are assessed using a seven-item scale items measured using a 6-point likert scale which reports students' awareness of employability-related personal strengths and challenges alongside the perceived alignment of their studies to their future career (Bennett & Ananthram, 2022). In the developing employability survey, career exploration and awareness were measured using (Lent et al., 2016) career exploration and decision self-efficacy scale (CEDSE). The scale was selected because of its relevance to making informed decisions. The developing employability survey asked students to respond to eight items measured using a 10-point Likert scale for career exploration and awareness. Career commitment relates to the extent to which students identify with and commit to their chosen study pathway and their career identity. This was measured in the developing survey using an eight-item scale developed by (Mancini et al., 2015) as part of their professional identity status questionnaire. The survey asked students to respond to eight items measured using a 5-point Likert scale for career commitment. Goal directed behaviour, from a social cognitive perspective, argues that learners' employability and career development is underpinned by their ability to operate as self-regulated learners. Expressed as goal-directed behaviour, the scale was derived from Coetzee (2014) and asked students to respond in the developing employability survey to 10 items measured using a 6-point likert scale.

**2.3 Analytic Strategy**

Specifically, we are interested to know if any changes in students' career confidence occurred due to COVID-19, or over the period 2018, 2019, 2020 and 2021. To address the research questions, the results from 2018, 2019, 2020 and 2021 were analysed using SPSS to understand 1) if students had any change in their reported outcomes for each item from each year, using a basic description of mean to achieve this, and 2) compare if students had any chance in their items score across and between the years. Using ANOVA and Kruskal-Wallis test with different sensitivity and variance standard we analysed each years data set to present mean and mean rank. Kruskal-Wallis test was used to compare multiple groups with orderly and quantitative data. Only where a difference was observed as the Kruskal-Wallis results presented.

**3 RESULTS**

Using the results obtained from 2018, 2019, 2020 and 2020 below we will present the career confidence of IT students. Along with students responding to the questions on the developing employability survey, demographic information of age and gender was also recorded.

**3.1 Age and Gender of Survey Respondents**

Using descriptive statistics, we found the mean age of students in our sample (n = 1349) was 23, the maximum age was 59. There was also no significant difference in age when comparing across years. This may be used to indicate similarity in the samples drawn from IT students in 2018, 2019, 2020 and 2021. While no difference was found between groups, comparison of the results on one demographic factor should lead the results to be taken with caution. Beyond age and in relation to students' career confidence, the following scales of: self-awareness related to program of study, career commitment, career exploration and awareness and goal directed behaviour, are reported to help gauge students career confidence and identity prior to and during the COVID 19 period. Table 1 shows age and gender as reported for each year the students completed the survey.

Table 1: Gender and Age of IT students in this study

|  |  | Female | Gender variant/ nonconforming | Male | Non-binary | Other | Transgender | Total |
|---|---|---|---|---|---|---|---|---|
| **Year** | 2018 | 68 | 0 | 298 | 3 | 2 | 1 | 372 |
|  | 2019 | 49 | 1 | 323 | 0 | 0 | 0 | 373 |
|  | 2020 | 75 | 2 | 311 | 1 | 0 | 0 | 389 |
|  | 2021 | 44 | 1 | 170 | 0 | 0 | 0 | 215 |
| **Total** |  | 236 | 4 | 1102 | 4 | 2 | 1 | 1349 |

**3.2 Career Confidence Survey Responses**

*3.2.1 Self-awareness related to program of study*

The results for IT student's self-awareness related to program of study showed through descriptive statistics that there was an observed difference in students self-awareness related to program of study, particular with identifying their personal weaknesses ("I can identify personal weaknesses in need of further development") between groups. When considering what year may have contributed to the observed difference, for self-awareness related to program of study the change in mean-rank for the 2021 group is observed, see table 2.

Table 2. Mean-rank by year for self-awareness related to program of study.

| I can identify personal weaknesses in need of further development | 2018 | 372 | 622.97 |
|---|---|---|---|
| | 2019 | 373 | 673.78 |
| | 2020 | 389 | 690.21 |
| | 2021 | 215 | 739.61 |
| | Total | 1349 | |

When using a Kruskal-Wallis test for self-awareness related to program or study, significant differences were observed between groups, with observed differences in students identifying their personal weaknesses ("I can identify personal weaknesses in need of further development") as shown in table 4. The results highlight the changes in students identifying personal weaknesses during COVID-19.

*3.2.2 Career commitment*

In relation to career commitment there was an observed significant difference between groups for IT students, specifically in relation to feeling like a professional in the discipline ("Does thinking of yourself as a professional in your discipline make you feel secure in your life?") along with feeling self-confident as a professional ("Does thinking of yourself as a professional in your discipline make you feel self-confident?). There was also an observed difference in career commitment in relation to students changing their major ("Are you considering the possibility of changing your university major in order to be able to practice another profession in the future?") between groups. When considering what year may have contributed to the observed difference, for career commitment the change in mean-rank for the 2021 group is observed, see table 3.

Table 3. Mean-rank by year for career commitment.

| Does thinking of yourself as a professional in your discipline make you feel secure in your life? | 2018 | 372 | 617.42 |
|---|---|---|---|
| | 2019 | 373 | 683.94 |
| | 2020 | 389 | 681.16 |
| | 2021 | 215 | 747.97 |
| | Total | 1349 | |
| Does thinking of yourself as a professional in your discipline make you feel self-confident? | 2018 | 372 | 631.08 |
| | 2019 | 373 | 682.10 |
| | 2020 | 389 | 685.80 |
| | 2021 | 215 | 719.13 |
| | Total | 1349 | |
| Are you considering the possibility of changing your university major in order to be able to practice another profession in the future? | 2018 | 372 | 719.32 |
| | 2019 | 373 | 685.43 |
| | 2020 | 389 | 650.05 |
| | 2021 | 215 | 625.37 |
| | Total | 1349 | |

When using a Kruskal-Wallis test for career commitment, a significant different was observed between groups, with observed difference in students feeling like a professional in the discipline, feeling self-confident as a professional, and possibility of changing major, as shown in table 4.

Table 4. Kruskal-Wallis H test results with observed difference for self-awareness and career commitment.

| | Kruskal-Wallis H | df | Asymp. Sig. |
|---|---|---|---|
| Self-Awareness related to program of study: *I can identify personal weaknesses in need of further development* | 14.635 | 3 | 0.002 |
| Career Commitment: *Does thinking of yourself as a professional in your discipline make you feel secure in your life?* | 17.818 | 3 | 0.000 |
| Career Commitment: *Does thinking of yourself as a professional in your discipline make you feel self-confident?* | 8.911 | 3 | 0.031 |
| Career Commitment: *Are you considering the possibility of changing your university major in order to be able to practice another profession in the future?* | 11.160 | 3 | 0.011 |

*3.2.3   Career exploration and awareness and goal directed behaviour*

The results for career exploration and awareness, and goal directed behaviour observed no significant difference between groups. Further there was no observed difference in the mean rank results across each year, with Kruskal-Wallis showing no significant difference observed between groups for students' career exploration and awareness. Thus, no further results are presented for these two factors.

**4   DISCUSSION**

The COVID-19 period saw major disruption to the student experience at university, with a shift to wholly online learning and remote professional practice experiences for most students. To help understand the implications of this time on students' career preparation, this paper has explored if there were any changes in the career confidence of students in IT during the 2018, 2019, 2020 and 2021 period. Specifically, this paper addressed the research questions: What is the career confidence of students in IT, leading up to and during the COVID-19 period? And, has COVID-19 changed IT students' career identity? This paper does not report IT students baseline career or employability data, rather compares changes across groups to understand students career confidence across the period. To understand IT students career confidence, this study reviewed their responses regarding self-awareness related to their program or study, career commitment, career exploration and awareness and goal directed behaviour, drawn from their completion of the developing employability survey in a second year IT unit. These factors were selected as they relate to the model of career development (Figure 1) that informs the career education initiatives at Deakin University, the context in which IT students were surveyed. As described in the model, to understand students career confidence and interest we need to explore their self-efficacy and outcome expectations. Efficacy beliefs and expectations can be related to student's self-awareness and their expectations as a professional. Further, career exploration and goal directed behaviour are incorporated into the model of career development through career interest and goals/choice.

The results showed that across the sample (n = 1349) there were no significant differences between groups when comparing gender, allowing comparison of career confidence across groups to occur. While no difference was found between groups, comparison of the results on one demographic factor should lead the results to be taken with caution. When considering the items that form self-awareness related to program of study, as they relate to students' career confidence, the results show IT students' self-awareness, specifically their ability to identify personal weaknesses in need of further development, changed in 2021 with the mean-rank noted to be significantly higher than previous years (table 3). While other items in self-awareness related to program of study did not observe any changes, the observed result may indicate students in 2021 had greater awareness of self-efficacy regarding their professional development needs. This result differs to other contexts, with students in accounting reporting lower perceived program relevance during their third year of study (Kercher et al., 2023).

In relation to career commitment, as it relates to career confidence, the results in this study show that in 2021 students were feeling more like a professional in the discipline, were more self-confident as a professional and reported less about changing their major, indicating commitment to their chosen profession. The items that observed a significant difference in relation to career commitment are shown in table 4. Specifically, there was an observed difference in IT students' responses regarding being a professional and how this impacted their notions of job security and self-confidence. Table 3 demonstrates mean rank of the results on career commitment, highlighting that in 2021 students felt more self-confident and more secure in their life to work in IT then previously. In addition, in 2021 students reported they were less likely to change their major. Again, this contrasts to accounting students who reported lower career commitment during their third year of study (Kercher et al., 2023). Career commitment relates to career identity, with students who show a strong career commitment more likely to build upon their forming professional identity (Holmes, 2001). While this study does not explore the study and employment context that impacts upon students forming career commitment, it is reported that nowadays students have higher career confidence due to concurrent work and study during their time at university (Brosnan et al., 2023). Yet, career related work when studying is not the norm (Brosnan et al., 2023). This can impact upon a student's ability to transfer skills and career development across contexts. In this study, anecdotal evidence suggests that the focus on digital transformation during COVID-19 may have impacted students' career commitment, strengthening their commitment to working in IT. Online and remote working, as an outcome of COVID-19, may not have had an impact upon professional expectations of working in IT.

When looking at career exploration and awareness, and goal directed behaviour, as they relate to students' career confidence, no significant difference was observed between groups during analysis. This result indicates that IT students in this study maintained their understanding and engagement in the profession, along with maintaining their career goals, during the COVID-19 period. Overall, the results indicate that prior and during the COVID-19 period IT students career confidence, specifically their career commitment and self-awareness maintained. In 2021 when COVID-19 restrictions became removed, the results indicate increased career commitment of IT students indicating their higher professional expectations to work in IT post-study along with greater self-awareness regarding their professional development needs.

Further, using the understanding of students' career confidence this study considered has COVID-19 changed IT students' career identity? Smith et al., (2019) argued that having a strong sense of career identity will ensure students make the most of career development opportunities while at university, with career identity impacting upon students ongoing career interest in IT, as shown in figure 1. Prior to COVID-19, IT students reported low commitment to their career identity, yet high academic confidence (Bennett et al., 2020). While career identity of IT students has not been directly captured in this study, the results presented show IT students exhibit greater self-awareness and self-confidence, indicating that career identity may not have been negatively impacted due to COVID-19. As identified in * et al., (2021) IT students rely on professional experiences (such as a placement or internship) to inform their career identity. This is backed-up by Smith et al., (2019) who found that university students were more confident in expressing aspects of professional identity when they engaged in experiential learning opportunities, such as work experience. It is interesting to note, that despite reduce placement opportunities and off-site professional experiences, IT students maintained their career identity during the COVID-19 period. More broadly, IT students' employability and professional prospects offer a degree of certainty based on their career confidence as observed in this study.

## 5 CONCLUSION

The projected growth in IT employment opportunities in Australia (Australian Government, 2023), along with an emphasis on digital skills, suggests many job prospects remain in this growing area. Yet COVID-19 brought on much disruption and uncertainty for working professionals, with the impact of this on those studying IT at university unknown. To help understand the impact of COVID-19 on the career confidence and career identity of university students, this study compared the results from n = 1349 IT students at an Australian University to understand if there were any changes in their career prospects. The results indicate that prior and during the COVID-19 period IT students career confidence, specifically their career commitment and self-awareness, maintained. In 2021 when COVID-19 restrictions began to be more flexible, the results indicate increased career commitment of IT students indicating they had increased their professional expectations to work in IT post-study along. In addition, the results show students had with greater self-awareness regarding their professional development needs in 2021. Despite reduced placement opportunities and off-site professional experiences, IT students maintained their career identity during the COVID-19 period.

Even with increased career confidence as observed in this study, supporting university students to explore their career options and build upon their career identity, and more broadly their employability, remains an important activity for universities to curate in their graduates. Employability support will translate not only to their success while at university, but to longer term career persistence, career confidence and labour market participation, ultimately resulting in "informed individuals, professionals and social citizens" (Bennett, 2016).

**ACKNOWLEDGMENTS**